\documentclass[12pt]{article}
\usepackage{graphicx}
\usepackage{amsmath}
\usepackage{amssymb}
\usepackage{caption2}
\begin{document}
\bibliographystyle{prsty}
\begin{center}
{\large {\bf \sc{   Analysis  of the  vertices $D^*D^* P$, $D^*D V$ and $DDV$ with light-cone QCD sum rules  }}} \\[2mm]
Zhi-Gang Wang \footnote{E-mail,wangzgyiti@yahoo.com.cn.  }    \\
Department of Physics, North China Electric Power University,
Baoding 071003, P. R. China
\end{center}

\begin{abstract}
In this article, we study the vertices $D^*D^*P$, $D^*DV$ and $DDV$
with the light-cone QCD sum rules.  The strong coupling constants
$g_{D^*D^*P}$, $g_{D^*DP}$, $f_{D^*DV}$, $f_{D^*D^*V}$,  $g_{DDV}$
and $g_{D^*D^*V}$  play an important role in understanding the
final-state interactions in the hadronic $B$ decays. They relate to
the basic parameters $g$, $\lambda$  and $\beta$  in the heavy quark
effective Lagrangian respectively. Our numerical values of the $g$,
$\beta$ and $\lambda$ are much smaller than  most of the existing
estimations. If the predictions from the light-cone QCD sum rules
are robust, the final-state interaction effects maybe overestimated
in the hadronic $B$ decays.
\end{abstract}


\section{Introduction}

Final-state interactions play an important role in the hadronic $B$
decays \cite{fsi,CHHY}. However, it is very difficult to take them
into account  in a systematic way due  to the nonperturbative nature
of the multi-particle dynamics. In practical calculations, we can
resort to phenomenological models to outcome the difficulty. The
one-particle-exchange model is typical (for example, one can consult
Ref.\cite{CHHY}), in this picture, the soft interactions  of the
intermediate states in  two-body channels with one-particle exchange
make the main contributions. The phenomenological Lagrangian has
many input parameters, which describe the strong couplings  among
the charmed mesons in the hadronic $B$ decays \cite{CHHY}. In the
following, we write down the relevant phenomenological Lagrangian
\cite{CHHY},
\begin{eqnarray}
\mathcal {L}&=&-ig_{D^*DP}\left(D^i \partial^\mu P_{ij}D^{*j}_\mu -
D^{*i}_\mu \partial^\mu P_{ij} D^j \right) \nonumber \\
&& -\frac{1}{2}g_{D^*D^*P}\epsilon_{\mu\nu\alpha\beta}D^{*\mu}_i
\partial^\nu P_{ij}\left(\overrightarrow{\partial^\alpha}-\overleftarrow{\partial^\alpha}\right)D^{*\beta}_j \nonumber \\
&&+ig_{DDV}D_i\left(\overrightarrow{\partial_\mu}-\overleftarrow{\partial_\mu}\right)D_j
V^{ \mu }_{ij} \nonumber\\
&&+2f_{D^*DV}\epsilon_{\mu\nu\alpha\beta}\partial^\mu V^{\nu}_{ij}
\left[D_i\left(\overrightarrow{\partial^\alpha}-\overleftarrow{\partial^\alpha}\right)D^{*\beta}_j
-D^{*\beta}_i\left(\overrightarrow{\partial^\alpha}-\overleftarrow{\partial^\alpha}\right)D_j
\right] \nonumber\\
&&+ig_{D^*D^*V}D^{*\nu}_i\left(\overrightarrow{\partial_\mu}-\overleftarrow{\partial_\mu}\right)D^{*}_{j\nu}
V^{ \mu }_{ij} \nonumber\\
&&+4if_{D^*D^*V}D^*_{i\mu}\left(\partial^\mu V_{ij}^\nu-\partial^\nu
V_{ij}^\mu\right)D^*_{j\nu}  \, , \\
D^*&=&(D^{*0},D^{*+},D^*_s) \, , \nonumber\\
D&=&(D^{0},D^{+},D_s) \, , \nonumber\\
P&=&\left(\begin{array}{ccc}
\frac{\pi^{0}}{\sqrt{2}}+\frac{\eta}{\sqrt{6}}&\pi^{+}&K^{+}\\
\pi^{-}&-\frac{\pi^{0}}{\sqrt{2}}+\frac{\eta}{\sqrt{6}}&
K^{0}\\
K^{-} &\bar{K}^{0}&-\sqrt{\frac{2}{3}}\eta
\end{array}\right) \, , \nonumber\\
V&=&\left(\begin{array}{ccc}
\frac{\rho^{0}}{\sqrt{2}}+\frac{\omega}{\sqrt{2}}&\rho^{+}&K^{*+}\\
\rho^{-}&-\frac{\rho^{0}}{\sqrt{2}}+\frac{\omega}{\sqrt{2}}&
K^{*0}\\
K^{*-} &\bar{K}^{*0}&\phi
\end{array}\right) \, ,
\end{eqnarray}
where we take the convention $\epsilon_{0123}=1$.

The strong coupling constants (for example, $g_{D^*DP}$,
$g_{D^*D^*P}$, etc.) can be estimated with the heavy quark effective
theory and chiral symmetry \cite{HQEFT}. In the heavy quark limit,
the strong coupling constants in the phenomenological Lagrangian can
be related to the basic parameters  $g$, $\lambda$ and $\beta$ in
the heavy quark effective Lagrangian (one can consult
Ref.\cite{HQEFT} for the heavy  quark effective Lagrangian and
relevant parameters, here we neglect them for simplicity.),

\begin{eqnarray}
g_{D^*D^*P}&=&\frac{g_{D^*DP}}{\sqrt{M_{D^*}M_D}}=\frac{2}{f_P}g \, ,\nonumber\\
 f_{D^*DV}&=&\frac{f_{D^*D^*V}}{M_{D^*}}=\frac{\lambda g_V}{\sqrt{2}}\, ,\nonumber\\
g_{DDV}&=&g_{D^*D^*V}=\frac{\beta g_V}{\sqrt{2}}\, ,
\end{eqnarray}
where  $g_V=5.8$  from the vector meson dominance theory
\cite{VMDgV}. For existing estimations  of the values of the $g$,
$\lambda$ and $\beta$, one can consult
Refs.\cite{VMD03,CasalbuoniLammbda,Colangelo97,Kim01,
 Khodjamirian99,Melikhov99,Becirevic99,Colangelo02,Stewart98,Wang06g,Wang0705}.

 In previous work \cite{Wang0705}, we study the strong coupling constants of the
 $DDV$ and $D^*DV$ with the light-cone QCD
 sum rules, the numerical values of the $g_{DDV}$ and $f_{D^*DV}$ are much smaller
 than the existing estimations
   based on  the vector meson dominance theory \cite{VMD03}.
  In this article,
  we study the strong coupling constants
 $g_{D^*D^*P}$, $f_{D^*D V}$, $g_{DD V}$   with the light-cone QCD
 sum rules \footnote{In
this article, we present the results for the strong coupling
constants $f_{D^*DV}$ and $g_{DDV}$ which are originally obtained in
Ref.\cite{Wang0705} explicitly, and perform a comprehensive analysis
of the strong coupling constants in Eq.(1). }. Furthermore, we
analyze the corresponding parameters $g$, $\lambda$, $\beta$
 in the heavy  quark
effective Lagrangian \cite{HQEFT}, and obtain the values of the
strong coupling constants $g_{D^*DP}$, $f_{D^*D^*V}$ and
$g_{D^*D^*V}$.

   The light-cone QCD sum rules carry
out   operator product expansion near the light-cone, $x^2\approx
0$, instead of  short distance, $x\approx 0$, while the
nonperturbative matrix elements are parameterized by the light-cone
distribution amplitudes (which are classified according to their
twists)  instead of
 the vacuum condensates \cite{LCSR,LCSRreview}.
 The nonperturbative
 parameters in the light-cone distribution amplitudes are calculated with   the conventional QCD  sum rules
 and the  values are universal \cite{SVZ79}.

The article is arranged as: in Section 2, we derive the strong
coupling constants  $g_{D^*D^* P}$, $f_{D^*D V}$ and $g_{DD V}$ with
the light-cone QCD sum rules; in Section 3, the numerical result and
discussion; and in Section 4, conclusion.

\section{Strong coupling constants  $g_{D^*D^* P}$, $f_{D^*D V}$ and $g_{DD V}$  with light-cone QCD sum rules}
 We study the
strong coupling constants $g_{D^*D^*P}$,   $f_{D^*DV}$ and $g_{DDV}$
with the two-point correlation functions $\Pi^{ij}_{\mu\nu}(p,q)$,
$\Pi^{ij}_\mu(p,q)$  and $\Pi^{ij}(p,q)$, respectively,
\begin{eqnarray}
\Pi^{ij}_{\mu\nu }(p,q)&=&i \int d^4x \, e^{-i q \cdot x} \,
\langle 0 |T\left\{J^i_\mu(0) {J^j_{\nu}}^+(x)\right\}|P_{ij}(p)\rangle \, , \\
\Pi^{ij}_{\mu}(p,q)&=&i \int d^4x \, e^{-i q \cdot x} \,
\langle 0 |T\left\{J^j_\mu(0) {J^j_{5}}^+(x)\right\}|V_{ij}(p)\rangle \, , \\
\Pi^{ij}(p,q)&=&i \int d^4x \, e^{-i q \cdot x} \,
\langle 0 |T\left\{J^i_{5}(0) {J^j_{5}}^+(x)\right\}|V_{ij}(p)\rangle \, , \\
J^i_\mu(x)&=&{\bar q}_i(x)\gamma_\mu   c(x)\, ,  \nonumber \\
J^i_5(x)&=&{\bar q}_i(x) i \gamma_5  c(x)\, ,
\end{eqnarray}
where the currents $J^i_\mu(x)$  and $J^i_5(x)$ interpolate the
mesons $D^{*0}$, $D^{*+}$, $D^*_s$, $D^{0}$, $D^{+}$  and $D_s$,
respectively. The $i$ denote the $u$, $d$ and $s$ quarks
respectively.  The external states $\pi$, $K$, $\rho$, $K^*$, $\phi$
have the four momentum $p_\mu$ with $p^2=m_\pi^2$, $m_{K}^2$,
$m_{\rho}^2$, $m_{K^*}^2$, $m_{\phi}^2$, respectively.

According to the basic assumption of current-hadron duality in the
QCD sum rules \cite{SVZ79}, we can insert  a complete series of
intermediate states with the same quantum numbers as the current
operators $J^i_\mu(x)$ and $J^i_5(x)$ into the correlation functions
$\Pi^{ij }_{\mu\nu}(p,q)$, $\Pi^{ij}_\mu(p,q)$  and $\Pi^{ij}(p,q)$
to obtain the hadronic representation. After isolating the ground
state contributions from the pole terms of the mesons $D^{*}_i$ and
$D_i$, we get the following results,
\begin{eqnarray}
\Pi^{ij}_{\mu\nu }(p,q) &=&\frac{f_{D^*_i}f_{D^*_j}
M_{D^*_i}M_{D^*_j}g_{D_i^*D^*_j P_{ij}}}
  {\left\{M_{D^*_i}^2-(q+p)^2\right\}\left\{M_{D^*_j}^2-q^2\right\}}\epsilon_{\mu\nu\alpha\beta}p^\alpha q^\beta   + \cdots  \nonumber\\
  &=&\Pi^1_{ij}(p,q) \epsilon_{\mu\nu\alpha\beta}p^\alpha q^\beta+ \cdots
  \,,\\
\Pi^{ij}_{\mu  }(p,q) &=&\frac{f_{D^*_i}f_{D_j}M_{D^*_i}M^2_{D_j}
f_{D_i^*D_j V_{ij}}}
  {(m_c+m_j)\left\{M_{D^*_i}^2-(q+p)^2\right\}\left\{M_{D_j}^2-q^2\right\}}4\epsilon_{\mu\nu\alpha\beta} \epsilon^\nu p^\alpha q^\beta    +
  \cdots \nonumber\\
  &=&\Pi^2_{ij}(p,q) \epsilon_{\mu\nu\alpha\beta}\epsilon^\nu p^\alpha q^\beta + \cdots
  \,,\\
\Pi^{ij }(p,q) &=&\frac{f_{D_i}f_{D_j}M_{D_i}^2M_{D_j}^2g_{D_iD_j
V_{ij}}}
{(m_i+m_c)(m_j+m_c)\left\{M_{D_i}^2-(q+p)^2\right\}\left\{M_{D_j}^2-q^2\right\}}2\epsilon
\cdot q    + \cdots    \nonumber\\
&=& \Pi^3_{ij}(p,q) \epsilon \cdot q    + \cdots ,
\end{eqnarray}
where the following definitions for the weak decay constants have
been used,
\begin{eqnarray}
\langle0 | J^i_5(0)|D_i(p)\rangle&=&\frac{f_{D_i}M_{D_i}^2}{m_i+m_c}\,, \nonumber\\
\langle0 |
J^i_\mu(0)|D^*_i(p)\rangle&=&f_{D^*_i}M_{D^*_i}\epsilon_\mu\,.
\end{eqnarray}
 In Eqs.(8-10), we have not shown the contributions from the high
resonances and continuum states explicitly as they are suppressed
due to the double Borel transformation. Non-conservation of the
vector currents $J^i_\mu(x)$ can lead to non-vanishing couplings
with the scalar mesons $D^0_0$, $D^+_0$ and $D_{s0}$,
\begin{eqnarray}
\langle0 | J^i_\mu(0)|D_{i0}(p)\rangle&=&f_{D_{i0}}p_\mu\,,
\end{eqnarray}
where the $f_{D_{i0}}$ are the weak decay constants. In this
article, we choose the tensor structure
$\epsilon_{\mu\nu\alpha\beta}p^\alpha q^\beta$ (or
$\epsilon_{\mu\nu\alpha\beta} \epsilon^\nu p^\alpha q^\beta$) for
analysis in Eqs.(8-9), the presence of the scalar mesons cannot
result in contaminations. We have alternative choice to use the
axial-vector currents $J^i_{5\mu}(0)$ to interpolate the
pseudoscalar mesons $D^0$, $D^+$ and $D_s$. However, the
axial-vector currents $J^i_{5\mu}(0)$ can also interpolate  the
axial-vector mesons $D_1^0$, $D_1^+$ and $D_{s1}$,
\begin{eqnarray}
\langle0|J^i_{5\mu}(0)|D_{i1}(p)\rangle&=&f_{D_{i1}}M_{D_{i1}}\epsilon_\mu\,,
\end{eqnarray}
where the $f_{D_{i1}}$ are the weak decay constants, we should be
careful to avoid  contaminations from the axial-vector mesons.

 In the following, we briefly outline
operator product expansion for the  correlation functions $\Pi^{ij
}_{\mu\nu}(p,q)$, $\Pi^{ij}_\mu(p,q)$ and $\Pi^{ij}(p,q)$ in
perturbative QCD theory. The calculations are performed at  large
spacelike momentum regions $(q+p)^2\ll 0$ and $q^2\ll 0$, which
correspond to small light-cone distance $x^2\approx 0$ required by
validity of the operator product expansion. We write down the
propagator of a massive quark in the external gluon field in the
Fock-Schwinger gauge firstly \cite{Belyaev94},
\begin{eqnarray}
&&\langle 0 | T \{q_i(x_1)\, \bar{q}_j(x_2)\}| 0 \rangle =
 i \int\frac{d^4k}{(2\pi)^4}e^{-ik(x_1-x_2)}\nonumber\\
 &&\left\{
\frac{\not\!k +m}{k^2-m^2} \delta_{ij} -\int\limits_0^1 dv\, g_s \,
G^{\mu\nu}_{ij}(vx_1+(1-v)x_2)
 \right. \nonumber \\
&&\left. \Big[ \frac12 \frac {\not\!k
+m}{(k^2-m^2)^2}\sigma_{\mu\nu} - \frac1{k^2-m^2}v(x_1-x_2)_\mu
\gamma_\nu \Big]\right\}\, .
\end{eqnarray}
The contributions proportional to $G_{\mu\nu}$ can give rise to
three-particle (and four-particle) meson distribution amplitudes
with a gluon (or quark-antiquark pair) in addition to the two
valence quarks, their corrections are usually not expected to play
any significant roles. For examples, in the decay $B \to
\chi_{c0}K$, the factorizable contribution is zero and the
non-factorizable contributions from the soft hadronic matrix
elements are too small to accommodate the experimental data
\cite{WangLH}; the net contributions from the three-valence particle
light-cone distribution amplitudes to the strong coupling constant
$g_{D_{s1}D^*K}$ are rather small, about $20\%$ \cite{Wang0611}. In
this article, we observe that the contributions from the
three-particle (quark-antiquark-gluon) light-cone distribution
amplitudes are less than $5\%$ for the strong coupling constants
$g_{D^*D^*P}$. The contributions of the three-particle
(quark-antiquark-gluon) distribution amplitudes of the mesons are
always of minor importance comparing with the two-particle
(quark-antiquark) distribution amplitudes in the light-cone QCD sum
rules.   In our previous work, we also study the four form-factors
$f_1(Q^2)$, $f_2(Q^2)$, $g_1(Q^2)$ and $g_2(Q^2)$ of the $\Sigma \to
n$ in the framework of the light-cone QCD sum rules approach up to
twist-6 three-quark light-cone distribution amplitudes and obtain
satisfactory results \cite{Wang06}.

In a word, we can neglect the contributions from the valence gluons
and make relatively rough estimations
 in the light-cone QCD sum rules. In this article, we take into
 account the three-particle light-cone distribution amplitudes  of the
 pseudoscalar mesons, and neglect the three-particle light-cone distribution amplitudes  of the
 vector mesons to avoid  cumbersome calculations.

 Substituting the above $c$ quark
propagator and the corresponding $\pi$, $K$, $\rho$, $K^*$ and
$\phi$ mesons light-cone distribution amplitudes into the
correlation functions $\Pi^{ij }_{\mu\nu}(p,q)$, $\Pi^{ij}_\mu(p,q)$
and $\Pi^{ij}(p,q)$, respectively,  and completing the integrals
over the variables $x$ and $k$, finally we obtain the analytical
results at the level of quark-gluon degrees of freedom. The explicit
expressions are presented in    appendix A.

In calculation, the two-particle $\pi$, $K$, $\rho$, $K^*$, $\phi$
mesons and three-particle $\pi$, $K$ mesons light-cone distribution
amplitudes have been used \cite{LCSR,Belyaev94,PSLC,VMLC}, the
explicit expressions are given in   appendixes B-C. The parameters
in the light-cone distribution amplitudes are scale dependent and
can be estimated with the QCD sum rules
\cite{LCSR,Belyaev94,PSLC,VMLC}. In this article, the energy scale
$\mu$ is chosen to be $\mu_c=\sqrt{M_D^2-m_c^2}\approx1 \rm{GeV}$.

Now we perform the double Borel transformation with respect to  the
variables $Q_1^2=-(p+q)^2$  and  $Q_2^2=-q^2$ for the correlation
functions $\Pi^{1}_{ij}$, $\Pi^{2}_{ij}$ and $\Pi^{3}_{ij}$ in
Eqs.(8-10), and obtain the analytical expressions of the invariant
functions in the hadronic representation,
\begin{eqnarray}
B_{M_2^2}B_{M_1^2}\Pi^1_{ij}&=&\frac{ g_{D_i^*D_j^*
P_{ij}}f_{D^*_i}f_{D^*_j} M_{D^*_i} M_{D^*_j}}{M_1^2M_2^2}
\exp\left[-\frac{M^2_{D_i^*}}{M_1^2}
-\frac{M^2_{D_j^*}}{M_2^2}\right] +\cdots, \nonumber\\
B_{M_2^2}B_{M_1^2}\Pi^2_{ij}&=&\frac{ 4f_{D_i^*D_j
V_{ij}}f_{D^*_i}f_{D_j}M_{D^*_i}M^2_{D_j} }{(m_c+m_j)M_1^2M_2^2}
\exp\left[-\frac{M^2_{D^*_i}}{M_1^2} -\frac{M^2_{D_j}}{M_2^2}\right]
+\cdots, \nonumber\\
B_{M_2^2}B_{M_1^2}\Pi^3_{ij}&=&\frac{ 2g_{D_iD_j
V_{ij}}f_{D_i}f_{D_j}M_{D_i}^2
M_{D_j}^2}{(m_c+m_i)(m_c+m_j)M_1^2M_2^2}
\exp\left[-\frac{M^2_{D_i}}{M_1^2} -\frac{M^2_{D_j}}{M_2^2}\right]
+\cdots,
\end{eqnarray}
where we have not shown  the contributions from the high resonances
and continuum states  explicitly for simplicity.

In order to match the duality regions below the thresholds $s^0_1$
and $s^0_2$ for the interpolating currents, we can express the
correlation functions $\Pi^{1}_{ij}$, $\Pi^{2}_{ij}$ and
$\Pi^{3}_{ij}$  at the level of quark-gluon degrees of freedom into
the following form,
\begin{eqnarray}
\Pi^a_{ij}&=& \int ds_1 \int ds_2 \frac{\rho^a_{ij}(s_1,s_2)}{
\left\{s_1-(q+p)^2\right\}\left\{s_2-q^2\right\}} \, ,
\end{eqnarray}
where the $\rho^a_{ij}(s_1,s_2)$ ($a=1,2,3$) are spectral densities,
then perform the double Borel transformation with respect to the
variables $Q_1^2$ and $Q_2^2$ directly. However, the analytical
expressions of the spectral densities $\rho^a_{ij}(s_1,s_2)$ are
hard to obtain, we have to resort to some approximations.  As the
contributions
 from the higher twist terms  are  suppressed by more powers of
 $\frac{1}{m_c^2-(q+up)^2}$ (or $\frac{1}{M^2}$), the net contributions of the  twist-3 and twist-4
  terms are of minor
importance, less  than  $10\%$ for the strong coupling constants
$g_{D^*D^*P}$, the continuum subtractions will not affect the
results remarkably (for the strong coupling constants $G_S$($D_{s0}
D_s^* \phi $) and $G_A$($ D_{s1}D_s \phi$), the contributions are
less than $20\%$ \cite{Wang07}.). The dominating contributions come
from the two-particle twist-2 terms  involving the $\phi(u)$,
$\phi_\perp(u)$ and $\phi_\parallel(u)$. We perform the same trick
as Refs.\cite{Belyaev94,Kim} and expand the amplitudes $\phi(u)$,
$\phi_\perp(u)$, $\phi_\parallel(u)$ and $\phi_\sigma(u)$ in terms
of polynomials of $1-u$, for example,
\begin{eqnarray}
\phi(u)&=&\sum_{k=0}^N b_k(1-u)^k=\sum_{k=0}^N b_k
\left(\frac{s_2-m_c^2}{s_2-q^2}\right)^k \,  ,
\end{eqnarray}
where the $b_k$  are  coefficients,  then introduce the variable
$s_1$  and the spectral densities are obtained.

After straightforward calculations, we obtain the final expressions
of the double Borel transformed correlation functions $\Pi^1_{ij}$,
$\Pi^2_{ij}$ and $\Pi^3_{ij}$  at the level of quark-gluon degrees
of freedom. The masses of  the charmed mesons are
$M_{D}=1.87\rm{GeV}$, $M_{D_s}=1.97\rm{GeV}$,
$M_{D^*}=2.010\rm{GeV}$ and $M_{D_s^*}=2.112\rm{GeV}$,
\begin{eqnarray}
\frac{M^2_{D^*}}{M^2_{D_s^*}}\approx0.91 \, , &&
\frac{M_{D}^2}{M^2_{D^*}}\approx0.87 \, , \nonumber\\
\frac{M^2_{D_s}}{M_{D^*}^2}\approx0.96 \, ,&&
\frac{M^2_{D_s}}{M^2_{D_s^*}}\approx0.87 \, , \nonumber\\
\frac{M^2_{D}}{M^2_{D_s}}\approx0.90 \, ,
&&\frac{M_{D}^2}{M^2_{D_s^*}}\approx0.78 \, ,
\end{eqnarray}
 there exist overlapping working windows for the two Borel
parameters $M_1^2$ and $M_2^2$, it is convenient to take the value
$M_1^2=M_2^2$,
\begin{eqnarray}
1=\frac{M_1^2}{M_2^2}&\approx&\frac{M^2_{D_i^*}}{M^2_{D_j^*}}\approx
\frac{M^2_{D^*_i}}{M^2_{D_j}}\approx \frac{M^2_{D_i}}{M^2_{D_j}}\,.
\end{eqnarray}

We introduce the threshold parameters $s_0=\rm{max}(s^0_1,s^0_2)$
($s_1^0$ and $s_2^0$ corresponding to $M_1^2$ and $M_2^2$,
respectively) and make the simple replacement,
\begin{eqnarray}
e^{-\frac{m_c^2+u_0(1-u_0)m_{P_{ij}}^2}{M^2}} &\rightarrow&
e^{-\frac{m_c^2+u_0(1-u_0)m_{P_{ij}}^2}{M^2}
}-e^{-\frac{s^0_{P_{ij}}}{M^2}}\, ,\nonumber \\
 e^{-\frac{m_c^2+u_0(1-u_0)m_{V_{ij}}^2}{M^2}} &\rightarrow&
e^{-\frac{m_c^2+u_0(1-u_0)m_{V_{ij}}^2}{M^2}
}-e^{-\frac{s^0_{V_{ij}}}{M^2}}
\end{eqnarray}
for the correlation functions $\Pi^1_{ij}$, $\Pi^2_{ij}$ and
$\Pi^3_{ij}$ respectively to subtract the contributions from the
high resonances and
  continuum states \cite{Belyaev94}.

  Finally we obtain the sum rules for the strong coupling
  constants $g_{D^*D^*P}$, $f_{D^*DV}$ and $g_{DDV}$,

\begin{eqnarray}
&&g_{D^*_iD_j^*P_{ij}}f_{D^*_i}f_{D_j^*}M_{D^*_i}M_{D_j^*} \exp\left\{-\frac{M_{D_i^*}^2}{M^2_1}-\frac{M^2_{D^*_j}}{M_2^2}\right\}\nonumber\\
&=& f_{P_{ij}} \left\{ M^2\phi(u_0)+\frac{ m_c
m_{P_{ij}}^2\phi_\sigma(u_0)}{3(m_i+m_j)}\right\}\nonumber\\
&&\left\{\exp\left[- \frac{m_c^2+u_0(1-u_0)m_{P_{ij}}^2}{M^2}
\right]-\exp\left[- \frac{s^0_{P_{ij}}}{M^2} \right] \right\} \nonumber\\
&&+f_{P_{ij}} m_{P_{ij}}^2\exp\left[-
\frac{m_c^2+u_0(1-u_0)m_{P_{ij}}^2}{M^2} \right]\left\{
-\frac{A(u_0)}{4} \left[1
+\frac{m_c^2}{M^2}\right] \right.\nonumber\\
&& -2  \int_0^{u_0} d\alpha_j \int_{u_0-\alpha_j}^{1-\alpha_j}
d\alpha_g
A_\parallel(1-\alpha_j-\alpha_g,\alpha_g,\alpha_j)\frac{\alpha_j+\alpha_g-u_0}{\alpha_g^2}\nonumber\\
&&-   \int_0^{u_0} d\alpha_j \int_{u_0-\alpha_j}^{1-\alpha_j}
d\alpha_g \frac{1}{\alpha_g}
\left[A_\perp-\frac{V_\parallel}{2}+\frac{V_\perp}{2}\right](1-\alpha_j-\alpha_g,\alpha_g,\alpha_j)\nonumber\\
&&+ \frac{d}{d u_0} \left[\int_0^{1-u_0} d\alpha_g
\int^{u_0}_{u_0-\alpha_g} d\alpha_j \int_0^{\alpha_j} d\alpha
+\int^1_{1-u_0} d\alpha_g \int^{1-\alpha_g}_{u_0-\alpha_g} d\alpha_j
\int_0^{\alpha_j} d\alpha\right] \nonumber\\
&&\frac{1}{\alpha_g} \left[
A_\perp+A_\parallel+\frac{V_\parallel}{2}+\frac{V_\perp}{2}\right](1-\alpha-\alpha_g,\alpha_g,\alpha) \nonumber\\
&&- \frac{d}{d u_0} \int_{1-u_0}^1 d\alpha_g \int_0^{\alpha_g}d\beta
\int_0^{1-\beta}d\alpha\frac{1-u_0}{\alpha_g^2} \nonumber\\
&&\left. \left[
A_\perp+A_\parallel+\frac{V_\parallel}{2}+\frac{V_\perp}{2}\right](1-\alpha-\beta,\beta,\alpha)\right\}
\ ,
\end{eqnarray}

\begin{eqnarray}
&&4f_{D^*_iD_j V_{ij}}\frac{f_{D^*_i} f_{D_j}M_{D^*_i}M^2_{D_j}}{m_c+m_j} \exp\left\{-\frac{M_{D^*_i}^2}{M^2_1}-\frac{M^2_{D_j}}{M_2^2}\right\}\nonumber\\
&=& f_{V_{ij}}^\perp  M^2\phi_\perp(u_0)\left\{\exp\left[-
\frac{m_c^2+u_0(1-u_0)m_{V_{ij}}^2}{M^2}
\right]-\exp\left[- \frac{s^0_{V_{ij}}}{M^2} \right] \right\} \nonumber\\
&&+\exp\left[- \frac{m_c^2+u_0(1-u_0)m_{V_{ij}}^2}{M^2}
\right]\left\{
\left[f_{V_{ij}}-f_{V_{ij}}^\perp\frac{m_i+m_j}{m_{V_{ij}}}\right]
\frac{m_c m_{V_{ij}}g_{\perp}^{(a)}(u_0)}{2}  \right.\nonumber\\
&&\left.-\frac{f_{V_{ij}}^\perp m_{V_{ij}}^2A_\perp(u_0)}{4}
\left[1 +\frac{m_c^2}{M^2}\right] \right\} \, ,
\end{eqnarray}

\begin{eqnarray}
&&2g_{D_iD_j V_{ij}}\frac{f_{D_i} f_{D_j}M^2_{D_i}M^2_{D_j}}{(m_c+m_i)(m_c+m_j)} \exp\left\{-\frac{M_{D_i}^2}{M^2_1}-\frac{M^2_{D_j}}{M_2^2}\right\}\nonumber\\
&=& f_{V_{ij}} m_{V_{ij}} M^2\phi_\parallel(u_0)\left\{\exp\left[-
\frac{m_c^2+u_0(1-u_0)m_{V_{ij}}^2}{M^2}
\right]-\exp\left[- \frac{s^0_{V_{ij}}}{M^2} \right] \right\} \nonumber\\
&&+\exp\left[- \frac{m_c^2+u_0(1-u_0)m_{V_{ij}}^2}{M^2}
\right]\left\{
\left[f_{V_{ij}}^\perp-f_{V_{ij}}\frac{m_i+m_j}{m_{V_{ij}}}\right]m_c
m_{V_{ij}}^2
 h_{||}^{(s)}(u_0) \right.\nonumber\\
&&\left.-\frac{f_{V_{ij}} m_{V_{ij}}^3A(u_0)}{4}  \left[1
+\frac{m_c^2}{M^2}\right] -2f_{V_{ij}} m_{V_{ij}}^3   \int_0^{u_0}
d\tau \int_0^\tau dt C(t)\left[1 +\frac{m_c^2}{M^2}\right]\right\}
\, ,
\end{eqnarray}

where
\begin{eqnarray}
u_0&=&\frac{M_1^2}{M_1^2+M_2^2}\, , \nonumber \\
M^2&=&\frac{M_1^2M_2^2}{M_1^2+M_2^2} \, .
\end{eqnarray}
Here we write down only the analytical results without the technical
details, one can consult  appendix D for some technical details.

In the following, we present another approach for subtracting  the
contributions from the high resonances and continuum states
\cite{Kim,LiHuang}. Firstly, we perform a double Borel
transformation with respect to the variables $Q_{1}^{2}$ and
$Q_{2}^{2}$ respectively,  and obtain the result,
\begin{eqnarray}
&&B_{M_2^2}B_{M_1^2}\int_0^1
du\frac{\Gamma(\alpha)f(u)}{\left\{m_c^2-(q+up)^2\right\}^\alpha}\nonumber\\
&=&\frac{M^{2(2-\alpha)}}{M_{1}^{2}M_{2}^{2}}\exp\left\{-\frac{m_c^2+u_0(1-u_0)p^2}
{M^{2}} \right\}f( u_{0}) \, ,\nonumber\\
&=&\frac{1}{M_{1}^{2}M_{2}^{2}}\int_{\Delta_1}^{s_1^0}ds_1
\int_{\Delta_2}^{s_2^0}ds_2 \exp\left\{-\frac{s_1}
{M_1^{2}}-\frac{s_2} {M_2^{2}} \right\} \rho(s_1,s_2) \, ,
\end{eqnarray}
where $f(u)$ stand for the light-cone distribution amplitudes,
$\Delta_1=\Delta_2=m_c^2+u_0(1-u_0)p^2$  and $\rho(s_1,s_2)$ stand
for the corresponding spectral densities.
 Then, we make a replacement $M_{1}^{2}\rightarrow \frac{1}{\sigma
_1}$, $M_{2}^{2}\rightarrow \frac{1}{\sigma _2}$ in above equation,
\begin{eqnarray}
&&\int_{\Delta_1}^{s_1^0}ds_1 \int_{\Delta_2}^{s_2^0}ds_2
 \exp\left\{-s_1\sigma_1-s_2\sigma_2 \right\}
\rho(s_1,s_2)\nonumber\\
 &=&\frac{f( u_{0})}{(\sigma_1
+\sigma_2)^{2-\alpha}}\exp\left\{-\left[m_c^2+u_0(1-u_0)p^2\right](\sigma_1+\sigma_2)\right\}\, ,\nonumber\\
&=&\frac{f( u_{0})}{\Gamma(2-\alpha)}\int_0^\infty d\lambda
\lambda^{1-\alpha}
\exp\left\{-\left[m_c^2+u_0(1-u_0)m_\pi^2+\lambda\right](\sigma_1+\sigma_2)\right\}
\,.
\end{eqnarray}
Finally, we take a double Borel transformation with respect to the
variables  $ \sigma _{1}$ and $ \sigma _{2}$ respectively, the
resulting QCD spectral density reads
\begin{eqnarray}
&&\int_{\Delta_1}^{s_1^0}ds_1 \int_{\Delta_2}^{s_2^0}ds_2
\exp\left\{-\frac{s_1} {M_1^{2}}-\frac{s_2} {M_2^{2}} \right\}
\rho(s_1,s_2)\nonumber\\
&=&\frac{f(u_0)}{\Gamma(2-\alpha)}\int_{\Delta}^{s_0}ds\left\{s-\left[m_c^2+u_0(1-u_0)p^2\right]\right\}^{1-\alpha}\exp\left\{-\frac{s}
{M^{2}} \right\}\, ,
\end{eqnarray}
i.e.
\begin{eqnarray}
&&B_{M_2^2}B_{M_1^2}\int_0^1
du\frac{f(u)}{\left\{m_c^2-(q+up)^2\right\}^\alpha}
\nonumber \\
&&=\frac{f(u_0)}{M_1^2M_2^2\Gamma(\alpha)\Gamma(2-\alpha)}\int_\Delta^{s_0}ds\left\{s-\left[m_c^2+u_0(1-u_0)p^2\right]\right\}^{1-\alpha}\exp\left\{-\frac{s} {M^{2}} \right\}\nonumber\\
&& +\cdots\, .
\end{eqnarray}
For the twist-2 terms, $\alpha=1$, the two subtracting approaches
lead to the same  results, the simple subtraction procedure
 we take  in Eq.(20) is still reasonable.

\section{Numerical result and discussion}

The input parameters are taken as $m_s=(0.14\pm 0.01 )\rm{GeV}$,
$m_c=(1.35\pm 0.10)\rm{GeV}$, $m_u=m_d=(0.0056\pm0.0016)\rm{GeV}$,
$f_K=0.160\rm{GeV}$, $f_\pi=0.130\rm{GeV}$,
$f_\rho=(0.216\pm0.003)\rm{GeV}$,
$f_\rho^{\perp}=(0.165\pm0.009)\rm{GeV}$,
$f_{K^*}=(0.220\pm0.005)\rm{GeV}$,
$f_{K^*}^{\perp}=(0.185\pm0.010)\rm{GeV}$,
$f_\phi=(0.215\pm0.005)\rm{GeV}$,
$f_\phi^{\perp}=(0.186\pm0.009)\rm{GeV}$ \cite{VMLC},
$m_K=0.498\rm{GeV}$, $m_{\pi} =0.138\rm{GeV}$,
$m_\rho=0.775\rm{GeV}$, $m_{K^*}=0.892\rm{GeV}$,
$m_\phi=1.02\rm{GeV}$ , $M_{D}=1.87\rm{GeV}$,
$M_{D_s}=1.97\rm{GeV}$, $M_{D^*}=2.010\rm{GeV}$ and
$M_{D_s^*}=2.112\rm{GeV}$.

For the $K$ meson: $\lambda_3=1.6\pm0.4$,
$f_{3K}=(0.45\pm0.15)\times 10^{-2}\rm{GeV}^2$,
$\omega_3=-1.2\pm0.7$, $\omega_4=0.2\pm0.1$, $a_2=0.25\pm 0.15$,
$a_1=0.06\pm 0.03$, $\eta_4=0.6\pm0.2$ \cite{LCSR,Belyaev94,PSLC}.

For the $\pi$ meson: $\lambda_3=0.0$, $f_{3\pi}=(0.45\pm0.15)\times
10^{-2}\rm{GeV}^2$, $\omega_3=-1.5\pm0.7$, $\omega_4=0.2\pm0.1$,
$a_2=0.25\pm 0.15$, $a_1=0.0 $, $\eta_4=10.0\pm3.0$
\cite{LCSR,Belyaev94,PSLC}.

For the $\rho$ meson: $a_1^{\parallel}=0.0$, $a_1^{\perp}=0.0$,
$a_2^{\parallel}=0.15\pm0.07$,  $a_2^{\perp}=0.14\pm0.06$,
$\zeta^{\parallel}_3=0.030\pm 0.010$,
$\widetilde{\lambda}_3^{\parallel}=0.0$,
$\widetilde{\omega}_3^{\parallel}=-0.09\pm 0.03$,
$\kappa_3^{\parallel}=0.0$,  $\omega_3^\parallel=0.15\pm0.05$,
$\lambda_3^\parallel=0.0$,  $\kappa_3^\perp=0.0$,
$\omega_3^\perp=0.55\pm0.25$,  $\lambda_3^\perp=0.0$,
$\zeta_4=0.15\pm 0.10$, $\zeta_4^T=0.10\pm 0.05$ and
$\widetilde{\zeta}_4^T=-0.10\pm 0.05$ \cite{VMLC}.

For the $K^*$ meson: $a_1^{\parallel}=0.03\pm0.02$,
$a_1^{\perp}=0.04\pm0.03$, $a_2^{\parallel}=0.11\pm0.09$,
 $a_2^{\perp}=0.10\pm0.08$,
 $\zeta^{\parallel}_3=0.023\pm 0.008$,
$\widetilde{\lambda}_3^{\parallel}=0.035\pm0.015$,
$\widetilde{\omega}_3^{\parallel}=-0.07\pm 0.03$,
$\kappa_3^{\parallel}=0.000\pm0.001$,
$\omega_3^\parallel=0.10\pm0.04$,
$\lambda_3^\parallel=-0.008\pm0.004$,
$\kappa_3^\perp=0.003\pm0.003$, $\omega_3^\perp=0.3\pm0.1$,
$\lambda_3^\perp=-0.025\pm 0.020$, $\zeta_4=0.15\pm 0.10$,
$\zeta_4^T=0.10\pm 0.05$ and $\widetilde{\zeta}_4^T=-0.10\pm 0.05$
\cite{VMLC}.

For the $\phi$ meson: $a_1^{\parallel}=0.0$, $a_1^{\perp}=0.0$,
$a_2^{\parallel}=0.18\pm0.08$, $a_2^{\perp}=0.14\pm0.07$,
$\zeta^{\parallel}_3=0.024\pm 0.008$,
$\widetilde{\lambda}_3^{\parallel}=0.0$,
$\widetilde{\omega}_3^{\parallel}=-0.045\pm 0.015$,
$\kappa_3^{\parallel}=0.0$, $\omega_3^\parallel=0.09\pm0.03$,
$\lambda_3^\parallel=0.0$, $\kappa_3^\perp=0.0$,
$\omega_3^\perp=0.20\pm0.08$, $\lambda_3^\perp=0.0$,
$\zeta_4=0.15\pm 0.10$, $\zeta_4^T=0.10\pm 0.05$ and
$\widetilde{\zeta}_4^T=-0.10\pm 0.05$ \cite{VMLC}.

The values of the decay constants $f_D$, $f_{D_s}$, $f_{D^*}$ and
$f_{D^*_s}$ vary in a large range from different approaches, for
example, the potential model, QCD sum rules and lattice QCD, etc
\cite{decayC1,decayC2}. For the decay constant $f_D$, we take the
experimental data from the CLEO Collaboration,
$f_D=(0.223\pm0.017)\rm{GeV}$  \cite{decayCP}. If we take the value
$f_{D_s}=(0.274\pm0.013)\rm{GeV}$ from the CLEO Collaboration, the
$SU(3)$ breaking effect is rather large, $\frac{f_{D_s}}{f_D}=1.23$,
while most of the theoretical estimations indicate
$\frac{f_{D_s}}{f_D}\approx 1.1$. In this article, we take the value
$\frac{f_{D_s}}{f_D} = 1.1$. For the decay constants $f_{D^*}$ and
$f_{D^*_s}$,  we take the
 central values  from lattice simulation \cite{decayCV},
$f_{D^*}=(0.23\pm0.02)\rm{GeV}$ and
$f_{D^*_s}=(0.25\pm0.02)\rm{GeV}$,
\begin{eqnarray}
\frac{f_{D^*_s}}{f_{D^*}}&\approx & \frac{f_{D_s}}{f_D} = 1.1 \, .
\end{eqnarray}

\begin{table}
\begin{center}
\begin{tabular}{|c|c|c|c|c|}
\hline\hline
   & $M_{\rm{gr}}(\rm{GeV})$ & $M_{\rm{gr}}(\rm{GeV})(\rm{exp})$ & $\sqrt{s_0}(\rm{GeV})$ &  $M_{\rm{2S}}(\rm{GeV})$  \\
\hline $0^-(c\bar{n})$  & 1.90 $\pm$ 0.03 & 1.869 & 2.45 $\pm$ 0.15
& 2.589 \\ \hline $1^-(c\bar{n})$  & 2.00 $\pm$ 0.02 & 2.010 & 2.55
$\pm$ 0.05 & 2.692 \\ \hline $0^-(c\bar{s})$  & 1.94 $\pm$ 0.03 &
1.969 & 2.50 $\pm$ 0.20 & 2.700 \\ \hline
$1^-(c\bar{s})$  & 2.05 $\pm$ 0.04 & 2.112 & 2.65 $\pm$ 0.15 & 2.806 \\
\hline\hline
\end{tabular}
\end{center}
\caption{ Numerical values of the ground state  masses $M_{\rm{gr}}$
and the threshold parameters  $\sqrt{s_{0}}$  from the QCD sum rules
\cite{Threshold}. We denote the first excited state as $2S$ state,
the values of the masses of the $2S$ states are taken from the
predictions of the quark model \cite{MassQM}. }
\end{table}

\begin{table}
\begin{center}
\begin{tabular}{|c|c|c|c|}
\hline\hline
    & $g_{D^*D^*P}$ &$g_{DDV}$  & $f_{D^*DV}$ \\ \hline
$ s^0_{\pi}(\rm{GeV}^2)$  & $6.5\pm 0.5$  &  &  \\ \hline $
s^0_{K^*}(\rm{GeV}^2)$  & $7.0\pm 0.5$  &  &  \\ \hline $
s^0_{\rho}(\rm{GeV}^2)$  &   &$6.0\pm 0.5$  & $6.5\pm 0.5$ \\ \hline
$ s^0_{K^*}(\rm{GeV}^2)$    & &$6.3\pm 0.5$  & $7.0\pm 0.5$ \\
\hline $ s^0_{\phi}(\rm{GeV}^2)$   &  &$6.3\pm 0.5$  & $7.0\pm 0.5$
\\       \hline  \hline
\end{tabular}
\end{center}
\caption{ Threshold parameters for the strong coupling constants
$g_{D^*D^*P}$, $g_{DDV}$  and $f_{D^*DV}$, respectively. }
\end{table}

 The duality threshold parameters  $s_0$ are shown in Table.2, the numerical (central) values of $s_0$ are
taken from the QCD sum rules for the masses of the pseudoscalar
mesons $D^0$, $D^+$, $D_s$ and vector mesons $D^{*0}$, $D^{*+}$,
$D^*_s$, see Table.1 \cite{Threshold}. The threshold parameters
$s^0_1$ and $s^0_2$ (corresponding to $M_1^2$ and $M_2^2$
respectively) are not equal in some channels in Eqs.(21-23), we
choose the larger one i.e. $s_0=\rm{max}(s_1^0,s_2^0)$  to take into
account all the contributions from the ground states, certainly,
there maybe some contaminations from the $2S$ state in the channel
with smaller threshold parameter i.e. $\rm{min}(s_1^0,s_2^0)$, and
impair the predictive power.
 The uncertainties of the threshold parameters  $s_0$ are about $\delta s_0=(0.25-1.0)\rm{GeV}^2$, see Table.1, in this article,
we take $\delta s_0=0.5\rm{GeV}^2$ for simplicity.

The Borel parameters are chosen as $ M_1^2=M_2^2$ and
$M^2=(3-7)\rm{GeV}^2$, in those regions, the values of the strong
coupling constants $g_{D^*D^*P}$, $g_{DDV}$ and $f_{D^*D V}$ are
rather stable.

In the limit of large Borel parameter $M^2$, the strong coupling
constants $g_{D^*D^*P}$, $f_{D^*DV}$ and $g_{DDV}$ take up the
following behaviors,
\begin{eqnarray}
g_{D^*_iD_j^*P_{ij}} &\propto& \frac{M^2\phi(u_0)}{f_{D^*_i}f_{D_j^*}}\propto \frac{M^2a_2}{f_{D^*_i}f_{D_j^*}}\, ,\nonumber\\
f_{D^*_iD_jV_{ij}}&\propto& \frac{M^2f_{V_{ij}}^\perp \phi_\perp(u_0)}{f_{D^*_i}f_{D_j}}\propto \frac{M^2f_{V_{ij}}^\perp a_2^\perp}{f_{D^*_i}f_{D_j}}\, ,\nonumber\\
g_{D_iD_jV_{ij}} &\propto&  \frac{M^2f_{V_{ij}}
\phi_\parallel(u_0)}{f_{D_i}f_{D_j}}\propto  \frac{M^2f_{V_{ij}}
a^\parallel_2}{f_{D_i}f_{D_j}}\, .
\end{eqnarray}
It is not unexpected, the contributions  from the twist-2 light-cone
distribution amplitudes $\phi(u)$, $\phi_\parallel(u)$ and
$\phi_\perp(u)$ are greatly enhanced by the large Borel parameter
$M^2$,  uncertainties of the relevant parameters presented in above
equations have significant impact on the numerical results.

Taking into account all the uncertainties, finally we obtain the
numerical values of   the strong coupling constants $g_{D^*D^*P}$,
$f_{D^*DV}$ and $g_{DDV}$, which are shown in Figs.1-3,
respectively,
\begin{eqnarray}
  g_{D^*D^*\pi} &=&(3.30 \pm 1.55)\rm{GeV}^{-1} \, , \nonumber \\
g_{D^*D^*_sK} &=&(3.50 \pm 1.57) \rm{GeV}^{-1}\, ,\nonumber \\
f_{D^*D\rho} &=&(0.89 \pm 0.15) \rm{GeV}^{-1}\, , \nonumber \\
f_{D^*D_sK^*} &=&(1.01 \pm 0.20) \rm{GeV}^{-1}\, , \nonumber \\
f_{D^*_sD_s\phi} &=&(0.82 \pm 0.16)\rm{GeV}^{-1} \, ,\nonumber \\
g_{DD\rho} &=&1.31 \pm 0.29 \, , \nonumber \\
g_{DD_sK^*} &=&1.61 \pm 0.32 \, , \nonumber \\
g_{D_sD_s\phi} &=&1.45 \pm 0.34 \, .
\end{eqnarray}
The average values are about
\begin{eqnarray}
  g_{D^*D^*P} &=&(3.40 \pm 1.55) \rm{GeV}^{-1}\, ,\nonumber \\
 f_{D^*DV} &=&(0.91 \pm 0.17) \rm{GeV}^{-1} \, ,\nonumber \\
 g_{DDV} &=&1.46 \pm 0.32\, .
\end{eqnarray}
The corresponding values  of the basic parameters $g$, $\lambda$ and
$\beta$ in the heavy quark effective Lagrangian can be obtained from
Eq.(3), and listed in Tables.3-5. From Tables.3-5 or Eq.(3), we can
obtain the values of the strong coupling constants $g_{D^*DP}$,
$f_{D^*D^*V}$ and $g_{D^*D^*V}$,
\begin{eqnarray}
  g_{D^*DP} &=&6.73 \pm 3.07  \, ,\nonumber \\
 f_{D^*D^*V} &=&1.85 \pm 0.35  \, ,\nonumber \\
 g_{D^*D^*V} &=&1.46 \pm 0.32\, .
\end{eqnarray}
Taking  the replacements $g_{DD\rho}\rightarrow
\frac{g_{DD\rho}}{2}$ and $f_{D^*D\rho}\rightarrow
\frac{f_{D^*D\rho}}{4}$ in Eq.(1),  we can obtain the same
definitions for the strong coupling constants in Ref.\cite{LiHuang}.
Our numerical values $g_{DD\rho} =2.62 \pm 0.58$ and $f_{D^*D\rho}
=(3.56 \pm 0.60)\rm{GeV}^{-1}$ are compatible with the predictions
$g_{DD\rho} =3.81 \pm 0.88$ and $f_{D^*D\rho} =(4.17 \pm 1.04)
\rm{GeV}^{-1}$ in Ref.\cite{LiHuang}. The value $g_{D^*DP} =6.73 \pm
3.07$ obtained from the relation of heavy quark effective theory is
different from the one obtained with the light-cone QCD sum rules
$g_{D^*D\pi} =12.5 \pm 1.0$ \cite{Belyaev94}. In Ref.\cite{LiHuang},
the authors take much smaller values for the decay constants of the
charmed mesons than the present work. In Ref.\cite{Belyaev94}, the
authors take much smaller value of the decay constant $f_{D}$ and
much larger value of the nonperturbative parameter $a_2(\mu)$ than
the present work. It is not unexpected that the numerical values are
different from each other, see Eq.(30). We can expect the relations
in Eq.(3) work well.

The values of the $g$ vary in a large range from different
approaches, see Table.3, the present prediction $g=0.22\pm0.10$ is
consistent with our previous calculation $g=0.16^{+0.07}_{-0.05}$
with the light-cone QCD sum rules \cite{Wang06g}. However, it is
much smaller than most of the existing estimations, this maybe due
to the shortcomings of the light-cone QCD sum rules.

The basic parameter $\lambda$ relates to the form-factor $V(q^2)$ of
the hadronic transitions $\langle V\mid \bar{q}
\gamma_\mu(1-\gamma_5)b \mid B \rangle $ and $\langle V\mid
\bar{q}\sigma_{\mu\nu}(1+\gamma_5)b \mid B \rangle $,
  which can be calculated with the
light-cone sum rules  approach and lattice QCD. With assumption of
the form-factor $V(q^2)$ at $q^2= q^2_{max}= (M_B-M_V)^2$ is
dominated by the nearest low-lying vector meson pole, we can obtain
the values  of the $\lambda$ \cite{VMD03,CasalbuoniLammbda}, which
are presented in Table.4. The parameter $\beta$ can be estimated
with the vector meson dominance theory, which is presented in
Table.5, for technical details, one consult Ref.\cite{Wang0705} .
The large discrepancies maybe that the vector meson dominance theory
overestimates
  the values of the $\beta g_V$ and
 $\lambda g_V$,
 the other possibility maybe
 the  shortcomings of the light-cone QCD sum rules.

\begin{table}
\begin{center}
\begin{tabular}{|c|c|}
\hline\hline
      $|g|$  &Reference  \\ \hline
      $0.38\pm0.08$ &\cite{HQEFT} \\      \hline
       $0.34\pm0.10$& \cite{Colangelo97} \\ \hline
 0.28& \cite{Kim01} \\ \hline
          $0.35\pm0.10$& \cite{Khodjamirian99} \\ \hline
  $0.50\pm0.02$& \cite{Melikhov99} \\ \hline
  $0.6\pm0.1$&\cite{Becirevic99} \\ \hline
         $0.59\pm0.07$& \cite{Colangelo02}  \\ \hline
   $0.27^{+0.06}_{-0.03}$& \cite{Stewart98} \\ \hline
 $0.16^{+0.07}_{-0.05}$  &  \cite{Wang06g}
\\ \hline
  $0.22\pm0.10$  &  This work
\\ \hline  \hline
\end{tabular}
\end{center}
\caption{ Numerical values of the  parameter $g$. }
\end{table}

\begin{table}
\begin{center}
\begin{tabular}{|c|c|}
\hline\hline
      $\beta$  &Reference  \\ \hline
            $0.9$& \cite{VMD03} \\ \hline
   $0.36\pm0.08$  &  This work
\\ \hline  \hline
\end{tabular}
\end{center}
\caption{ Numerical values of the  parameter $\beta$. }
\end{table}

\begin{table}
\begin{center}
\begin{tabular}{|c|c|}
\hline\hline
      $|\lambda|(\rm{GeV}^{-1})$  &Reference  \\ \hline
      $0.56$& \cite{VMD03} \\ \hline
      $0.63\pm0.17$ &\cite{CasalbuoniLammbda}\\      \hline
          $0.22\pm0.04$  &  This work
\\ \hline  \hline
\end{tabular}
\end{center}
\caption{ Numerical values of the  parameter $\lambda$. }
\end{table}

We can borrow some idea from the
  strong coupling constant $g_{D^*D\pi}$, the central value
($g_{D^*D\pi}=12.5$ or $g_{D^*D\pi}=10.5$ with the radiative
corrections are included in) from the light-cone QCD sum rules is
too small to take into account
  the value ($g_{D^*D\pi}=17.9$) from the
experimental data \cite{Khodjamirian99,Belyaev94,Becirevic03}.
Naively, we can expect that the contributions from the  radiative
corrections cannot smear the discrepancies between our predictions
and other estimations for the strong coupling constants
$g_{D^*D^*P}$, $f_{D^*DV}$ and $g_{DDV}$.  It has been noted that
the simple quark-hadron duality ansatz which works in the
one-variable dispersion relation might be too crude for the double
dispersion relation \cite{KhodjamirianConf}. As in
Ref.\cite{Becirevic03},  we can postpone the threshold parameters
$s_0$ to larger values to include the contributions from  the radial
excitations ($D'$ or ${D^{*}}'$) to the hadronic spectral densities,
with additional assumption for the values of the $g_{{D^*}'D^*P}$,
$g_{D'DV}$, $f_{{D^*}'DV}$, $f_{D^*D'V}$, etc,   we can improve the
values of the $g_{D^*D^*P}$, $g_{DDV}$ and $f_{D^*DV}$, and smear
the discrepancies between our values and other predictions. It is
somewhat of fine-tuning.

From Tables.3-5, we can see that our numerical values are much
smaller than most of the existing estimations (for example, the
values taken in Ref.\cite{CHHY}).
 Naively, we can expect that  smaller values of the strong
coupling constants  $g_{D^*D^*P}$, $g_{D^*DP}$,
 $f_{D^*DV}$,
$f_{D^*D^*V}$, $ g_{DDV} $ and $g_{D^*D^*V}$  lead to smaller
final-state interaction effects in the hadronic $B$ decays. For
example, the contributions from the rescattering mechanism in the
decay
\begin{eqnarray}
B\rightarrow D^* \rho \rightarrow D\pi \nonumber
\end{eqnarray}
can occur  through exchange of $D^*$ (or $D$) in the $t$ channel for
the sub-precess $D^* \rho \rightarrow D\pi$ \cite{CHHY}. The
amplitude of the rescattering Feynman diagrams  is proportional to
\begin{eqnarray}
C_1g_{D^*D^*\pi}f_{D^*D\rho}+C_2g_{D^*D\pi}g_{DD\rho}\propto D_1
g\lambda  +D_2 g\beta  \, ,
\end{eqnarray}
where the $C_i$ and $D_i$ are some  coefficients.

\begin{figure}
\centering
  \includegraphics[totalheight=6cm,width=7cm]{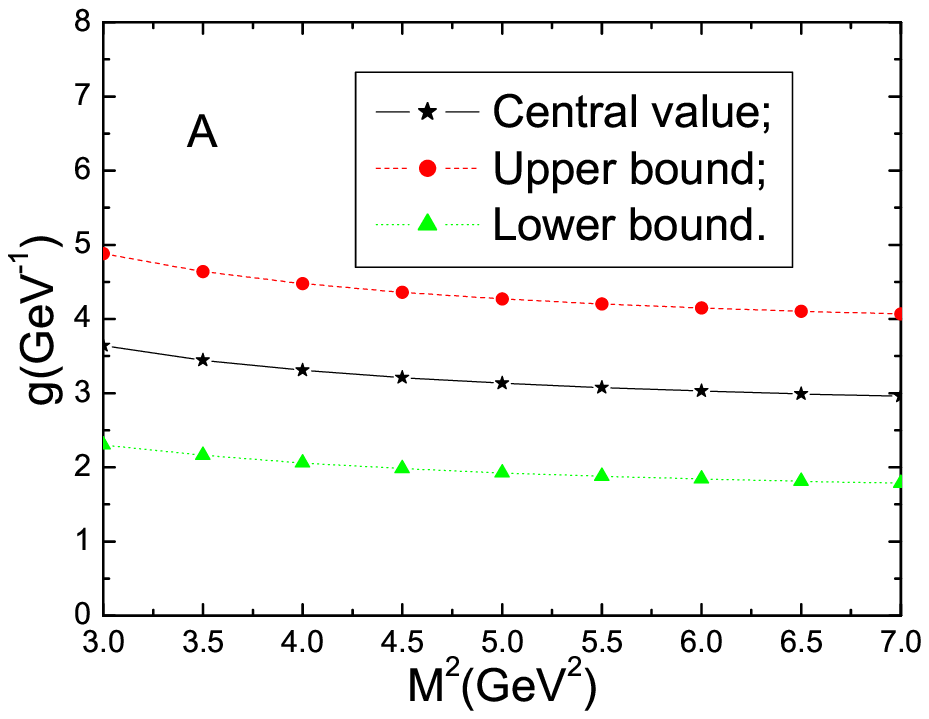}
  \includegraphics[totalheight=6cm,width=7cm]{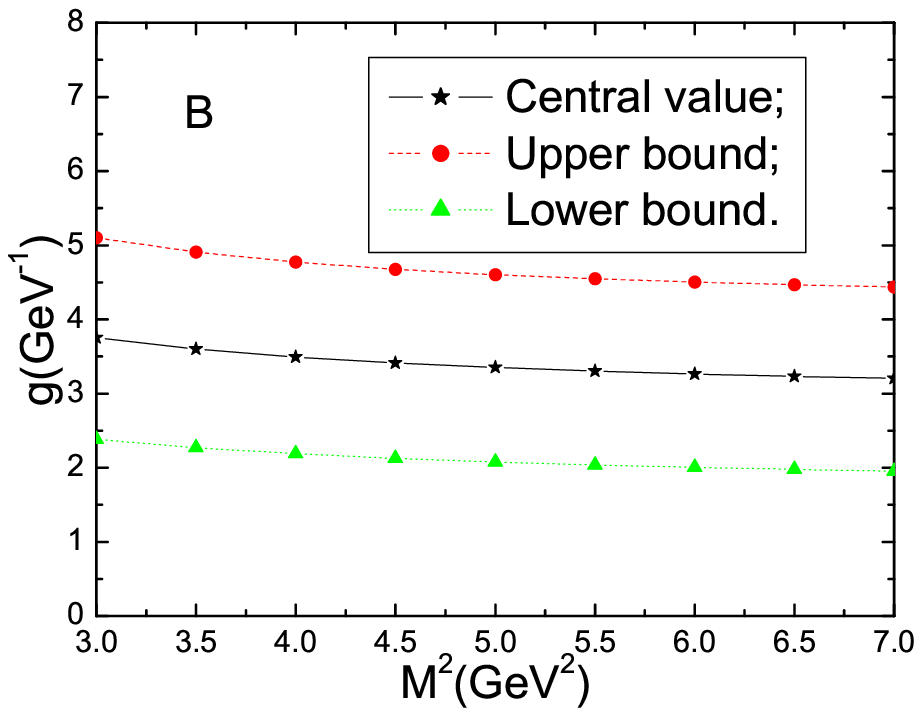}
         \caption{   $g_{D^*D^*\pi}$(A) and $g_{D^*D^*_sK}$(B) with the Borel parameter $M^2$ after taking into account all the uncertainties. }
\end{figure}

\begin{figure}
\centering
  \includegraphics[totalheight=6cm,width=7cm]{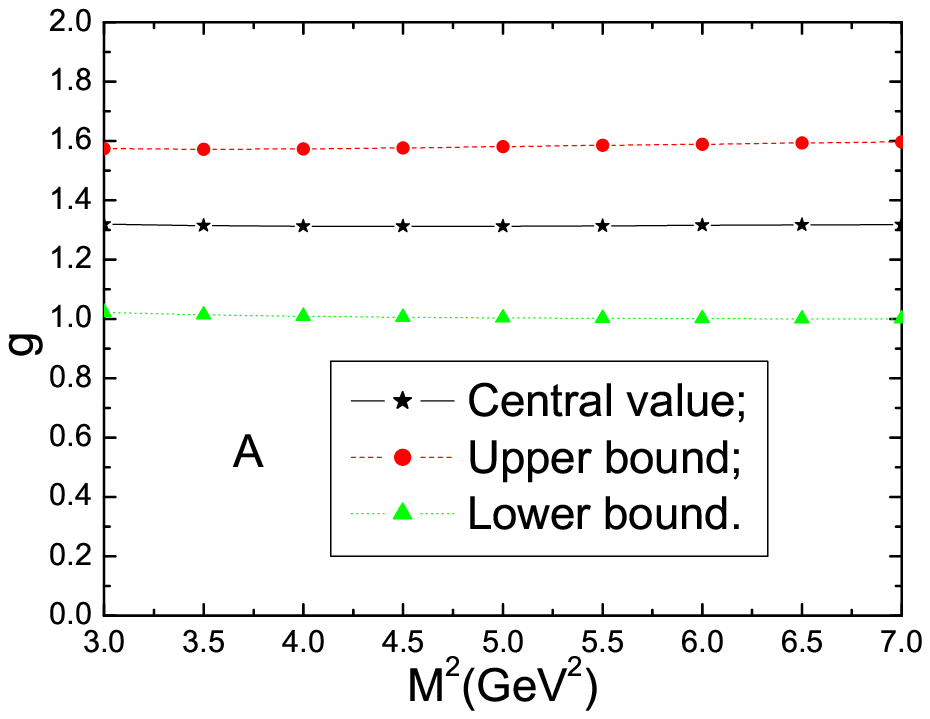}
  \includegraphics[totalheight=6cm,width=7cm]{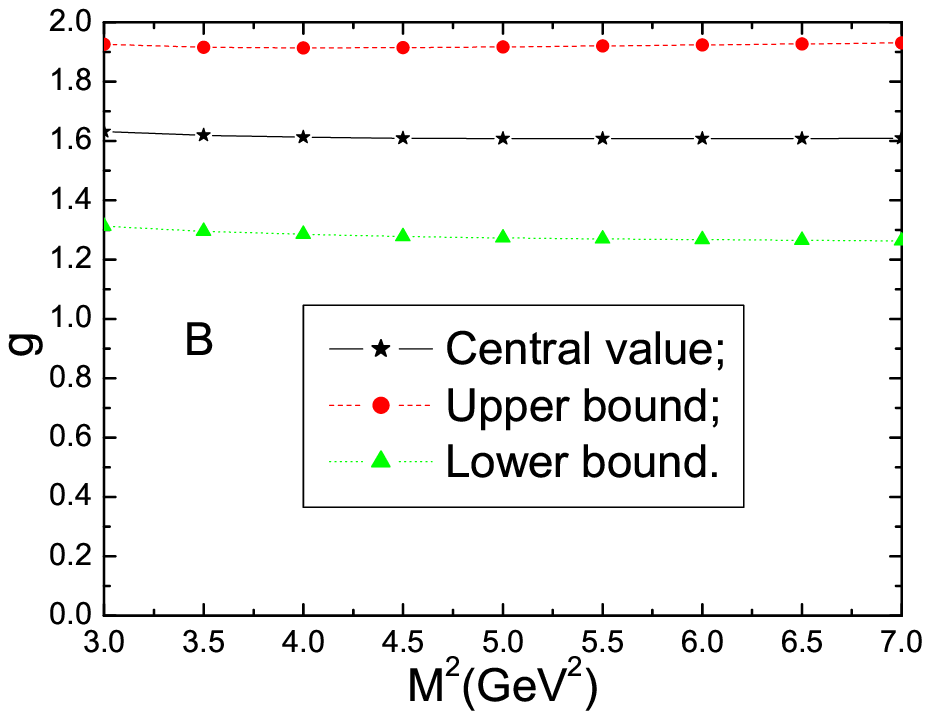}
\includegraphics[totalheight=6cm,width=7cm]{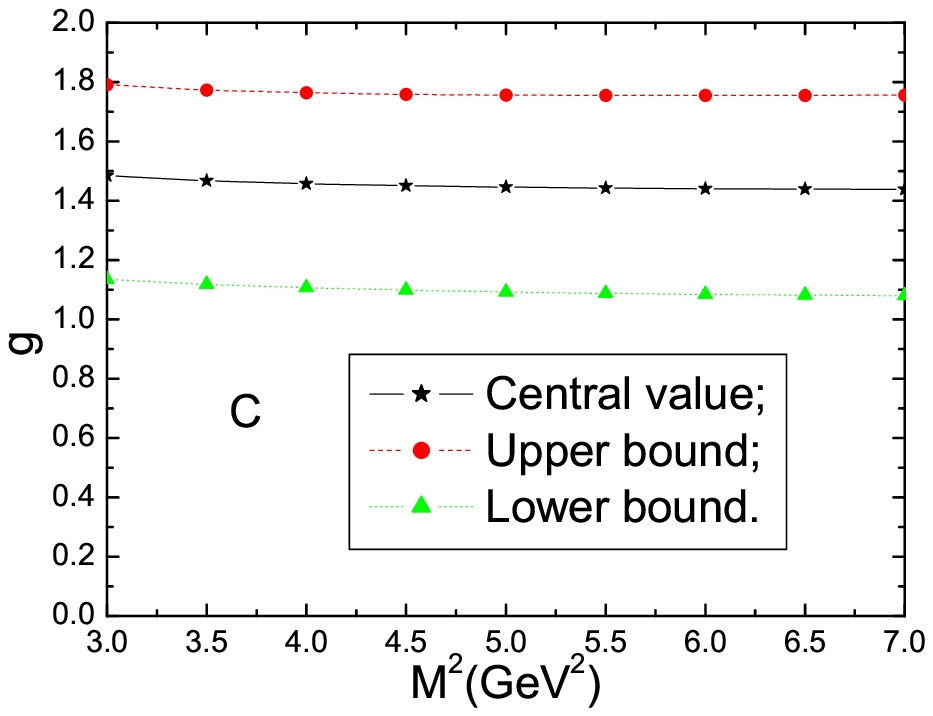}
         \caption{  $g_{DD\rho}$(A), $g_{DD_sK^*}$(B) and $g_{D_sD_s\phi}$(C)  with the Borel parameter $M^2$ after taking into account all the uncertainties. }
\end{figure}

\begin{figure}
\centering
  \includegraphics[totalheight=6cm,width=7cm]{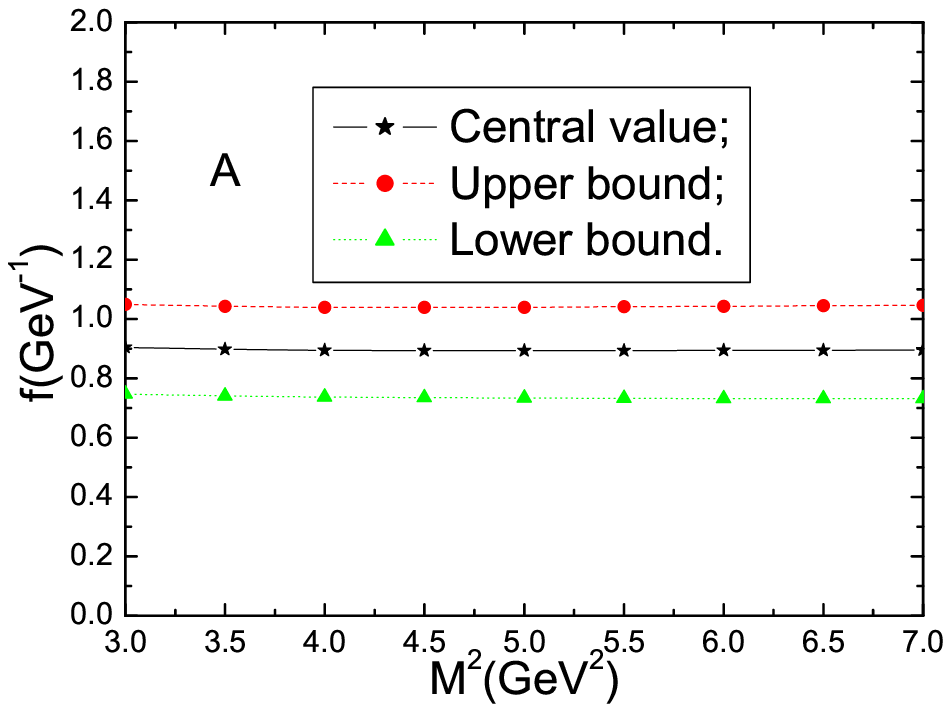}
  \includegraphics[totalheight=6cm,width=7cm]{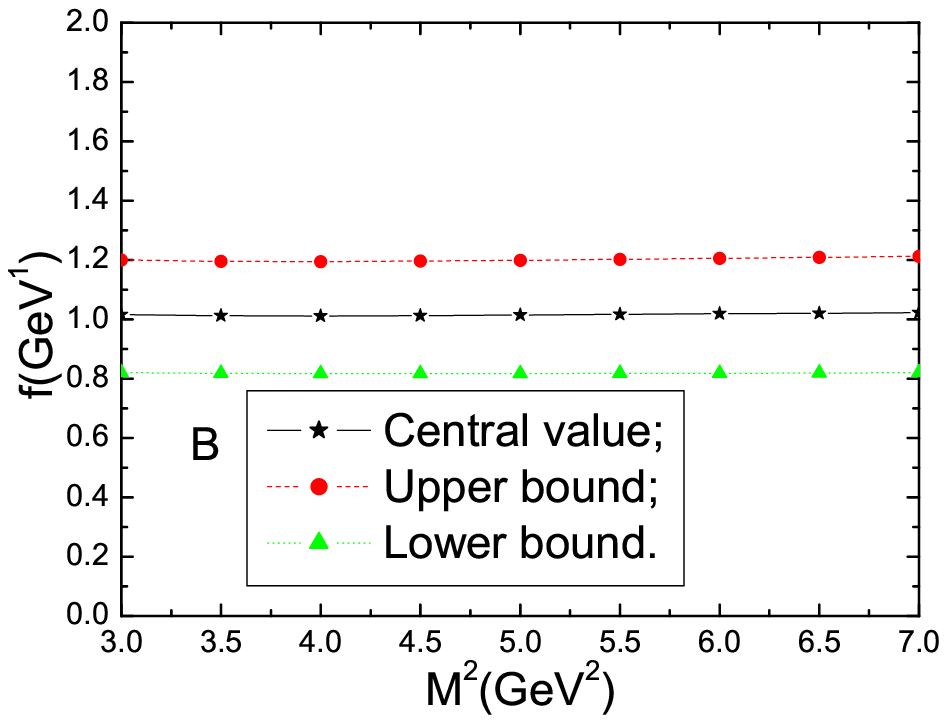}
\includegraphics[totalheight=6cm,width=7cm]{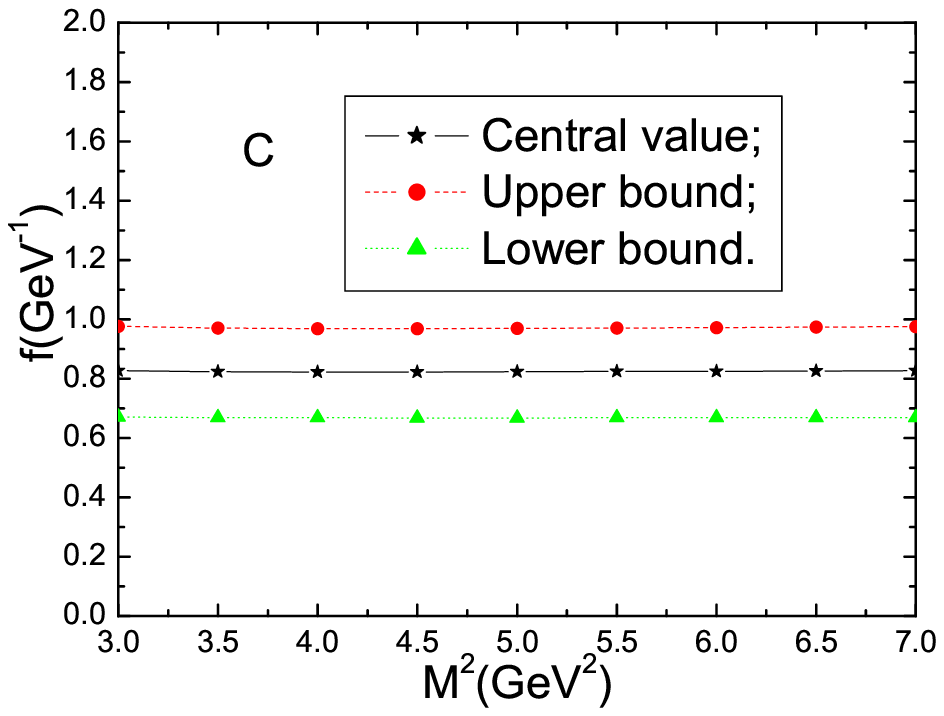}
         \caption{  $f_{D^*D\rho}$(A), $f_{D^*D_sK^*}$(B) and $f_{D^*_sD_s\phi}$(C)  with the Borel parameter $M^2$ after taking into account all the uncertainties. }
\end{figure}

\section{Conclusion}
In this article, we study the vertices $D^*D^*P$, $D^*DV$ and $DDV$
with the light-cone QCD sum rules.  The strong coupling constants
$g_{D^*D^*P}$, $g_{D^*DP}$, $f_{D^*DV}$, $f_{D^*D^*V}$,  $g_{DDV}$
and $g_{D^*D^*V}$ play an important role in understanding the
final-state interactions in the hadronic $B$ decays. They  relate to
the basic parameters $g$, $\lambda$ and $\beta$ respectively in the
heavy quark effective Lagrangian. Our numerical values of the $g$,
$\beta$ and $\lambda$ are much smaller than  most of the existing
estimations. If the predictions from the light-cone QCD sum rules
are robust, the final-state interaction effects maybe overestimated
in the hadronic $B$ decays.

 \section*{Appendix \, A}
The explicit expressions of the $\Pi^1_{ij}$, $\Pi^2_{ij}$ and
$\Pi^3_{ij}$  at the level of quark-gluon degrees of freedom,

\begin{eqnarray}
 \Pi^1_{ij} &=&f_{P_{ij}} \int_0^1 du \frac{\phi(u)}{AA}+\frac{f_{P_{ij}} m_c
m_{P_{ij}}^2}{3(m_i+m_j)} \int_0^1 du \frac{\phi_\sigma(u)
}{AA^2} \nonumber\\
&&-\frac{f_{P_{ij}} m_{P_{ij}}^2}{4} \int_0^1 du A(u) \left[
\frac{1}{AA^2}
+\frac{2m_c^2}{AA^3}\right]\nonumber \\
&&-f_{P_{ij}} m_{P_{ij}}^2 \int_0^1 d\alpha_g \int^{1-\alpha_g}_0
d\alpha_j \int_0^1 dv  \frac{1}{AA^2}\mid_{u=\alpha_j+(1-v)\alpha_g}
\nonumber\\
&& \left[
A_\perp+2vA_\parallel-\frac{V_\parallel}{2}+\frac{V_\perp}{2}\right](1-\alpha_j-\alpha_g,\alpha_g,\alpha_j) \nonumber\\
&& -f_{P_{ij}} m_{P_{ij}}^2 \int_0^1dv \int_0^1 d\alpha_g
\int_0^{1-\alpha_g}d\alpha_j \int_0^{\alpha_j} d\alpha
\frac{d}{du}\frac{1}{AA^2}\mid_{u=\alpha_j+(1-v)\alpha_g} \nonumber\\
&&\left[
A_\perp+A_\parallel+\frac{V_\parallel}{2}+\frac{V_\perp}{2}\right](1-\alpha-\alpha_g,\alpha_g,\alpha) \nonumber\\
&& +f_{P_{ij}} m_{P_{ij}}^2\int_0^1dvv \int_0^1 d\alpha_g
\int_0^{\alpha_g}d\beta \int_0^{1-\beta} d\alpha
 \frac{d}{du}\frac{1}{AA^2}\mid_{u=1-v\alpha_g} \nonumber\\
&&\left[
A_\perp+A_\parallel+\frac{V_\parallel}{2}+\frac{V_\perp}{2}\right](1-\alpha-\beta,\beta,\alpha)+\cdots
\, , \\
\Pi^2_{ij} &=&f_{V_{ij}}^\perp  \int_0^1 du
\frac{\phi_\perp(u)}{AA}-\frac{f_{V_{ij}}^\perp m_{V_{ij}}^2}{4}
\int_0^1 du A_\perp(u) \left[ \frac{1}{AA^2}
+\frac{2m_c^2}{AA^3}\right]\nonumber \\
&&+\left[f_{V_{ij}}-f_{V_{ij}}^\perp\frac{m_i+m_j}{m_{V_{ij}}}\right]\frac{m_cm_{V_{ij}}}{2}
\int_0^1 du \frac{g_{\perp}^{(a)}(u) }{AA^2} +\cdots\, ,
\\
\Pi^3_{ij} &=&f_{V_{ij}} m_{V_{ij}} \int_0^1 du
\frac{\phi_\parallel(u)}{AA}+\left[f_{V_{ij}}^\perp-f_{V_{ij}}\frac{m_i+m_j}{m_{V_{ij}}}\right]m_cm_{V_{ij}}^2
\int_0^1 du \frac{h_{||}^{(s)}(u)
}{AA^2} \nonumber\\
&&-\frac{f_{V_{ij}} m_{V_{ij}}^3}{4} \int_0^1 du A(u) \left[
\frac{1}{AA^2}
+\frac{2m_c^2}{AA^3}\right]\nonumber \\
&&-2f_{V_{ij}} m_{V_{ij}}^3 \int_0^1 du \int_0^u d\tau \int_0^\tau
dt C(t)\left[ \frac{1}{AA^2} +\frac{2m_c^2}{AA^3}\right] +\cdots\, ,
\end{eqnarray}
where
\begin{eqnarray}
AA&=&m_c^2-(q+u\,p)^2 \, .\nonumber
\end{eqnarray}

\section*{Appendix \, B}
 The light-cone distribution amplitudes of the $K$ meson are defined
 by
\begin{eqnarray}
\langle0| {\bar u} (0) \gamma_\mu \gamma_5 s(x) |K(p)\rangle& =& i
f_K p_\mu \int_0^1 du  e^{-i u p\cdot x}
\left\{\phi(u)+\frac{m_K^2x^2}{16}
A(u)\right\}\nonumber\\
&&+\frac{i}{2}f_K m_K^2\frac{x_\mu}{p\cdot x}
\int_0^1 du  e^{-i u p \cdot x} B(u) \, , \nonumber\\
\langle0| {\bar u} (0) i \gamma_5 s(x) |K(p)\rangle &=& \frac{f_K
m_K^2}{
m_s+m_u}\int_0^1 du  e^{-i u p \cdot x} \phi_p(u)  \, ,  \nonumber\\
\langle0| {\bar u} (0) \sigma_{\mu \nu} \gamma_5 s(x) |K(p)\rangle
&=&i(p_\mu x_\nu-p_\nu x_\mu)  \frac{f_K m_K^2}{6 (m_s+m_u)}
\int_0^1 du
e^{-i u p \cdot x} \phi_\sigma(u) \, ,  \nonumber\\
\langle0| {\bar u} (0) \sigma_{\mu \nu} \gamma_5 g_s G_{\alpha \beta
}(v x)s(x) |K(p)\rangle&=& f_{3 K}\left\{(p_\mu p_\alpha g^\bot_{\nu
\beta}-p_\nu p_\alpha g^\bot_{\mu \beta}) -(p_\mu p_\beta g^\bot_{\nu \alpha}\right.\nonumber\\
&&\left.-p_\nu p_\beta g^\bot_{\mu \alpha})\right\} \int {\cal
D}\alpha_i \phi_{3} (\alpha_i)
e^{-ip \cdot x(\alpha_s+v \alpha_g)} \, ,\nonumber\\
\langle0| {\bar u} (0) \gamma_{\mu} \gamma_5 g_s G_{\alpha
\beta}(vx)s(x) |K(p)\rangle&=&  f_Km_K^2p_\mu  \frac{p_\alpha
x_\beta-p_\beta x_\alpha}{p
\cdot x}\nonumber\\
&&\int{\cal D}\alpha_i A_{\parallel}(\alpha_i) e^{-ip\cdot
x(\alpha_s +v \alpha_g)}\nonumber \\
&&+ f_Km_K^2 (p_\beta g^\perp_{\alpha\mu}-p_\alpha
g^\perp_{\beta\mu})\nonumber\\
&&\int{\cal D}\alpha_i A_{\perp}(\alpha_i)
e^{-ip\cdot x(\alpha_s +v \alpha_g)} \, ,  \nonumber\\
\langle0| {\bar u} (0) \gamma_{\mu} i g_s \tilde G_{\alpha
\beta}(vx)s(x) |K(p)\rangle&=& f_Km_K^2 p_\mu  \frac{p_\alpha
x_\beta-p_\beta x_\alpha}{p \cdot
x}\nonumber\\
&&\int{\cal D}\alpha_i V_{\parallel}(\alpha_i) e^{-ip\cdot
x(\alpha_s +v \alpha_g)}\nonumber \\
&&+ f_Km_K^2 (p_\beta g^\perp_{\alpha\mu}-p_\alpha g^\perp_{\beta\mu})\nonumber\\
&&\int{\cal D}\alpha_i V_{\perp}(\alpha_i) e^{-ip\cdot x(\alpha_s +v
\alpha_g)} \, ,
\end{eqnarray}
where $g_{\mu\nu}^\perp=g_{\mu\nu}-\frac{p_\mu x_\nu+p_\nu x_\mu}{p
\cdot x}$, $\tilde G_{\mu \nu}= \frac{1}{2} \epsilon_{\mu\nu
\alpha\beta} G^{\alpha\beta} $ and ${\cal{D}} \alpha_i =d \alpha_u d
\alpha_s d \alpha_g \delta(1-\alpha_u -\alpha_s -\alpha_g)$.

The light-cone distribution amplitudes of the $K$ meson are
parameterized as
\begin{eqnarray}
\phi(u,\mu)&=&6u(1-u)
\left\{1+a_1C^{\frac{3}{2}}_1(\xi)+a_2C^{\frac{3}{2}}_2(\xi)
\right\}\, , \nonumber\\
\phi_p(u,\mu)&=&1+\left\{30\eta_3-\frac{5}{2}\rho^2\right\}C_2^{\frac{1}{2}}(\xi)\nonumber \\
&&+\left\{-3\eta_3\omega_3-\frac{27}{20}\rho^2-\frac{81}{10}\rho^2 a_2\right\}C_4^{\frac{1}{2}}(\xi)\, ,  \nonumber \\
\phi_\sigma(u,\mu)&=&6u(1-u)\left\{1
+\left[5\eta_3-\frac{1}{2}\eta_3\omega_3-\frac{7}{20}\rho^2-\frac{3}{5}\rho^2 a_2\right]C_2^{\frac{3}{2}}(\xi)\right\}\, , \nonumber \\
\phi_{3}(\alpha_i,\mu) &=& 360 \alpha_u \alpha_s \alpha_g^2 \left
\{1 +\lambda_3(\alpha_u-\alpha_s)+ \omega_3 \frac{1}{2} ( 7 \alpha_g
- 3) \right\} \, , \nonumber\\
V_{\parallel}(\alpha_i,\mu) &=& 120\alpha_u \alpha_s \alpha_g \left(
v_{00}+v_{10}(3\alpha_g-1)\right)\, ,
\nonumber \\
A_{\parallel}(\alpha_i,\mu) &=& 120 \alpha_u \alpha_s \alpha_g
a_{10} (\alpha_s-\alpha_u)\, ,
\nonumber\\
V_{\perp}(\alpha_i,\mu) &=& -30\alpha_g^2
\left\{h_{00}(1-\alpha_g)+h_{01}\left[\alpha_g(1-\alpha_g)-6\alpha_u
\alpha_s\right] \right.  \nonumber\\
&&\left. +h_{10}\left[
\alpha_g(1-\alpha_g)-\frac{3}{2}\left(\alpha_u^2+\alpha_s^2\right)\right]\right\}\,
, \nonumber\\
A_{\perp}(\alpha_i,\mu) &=&  30 \alpha_g^2 (\alpha_u-\alpha_s) \left\{h_{00}+h_{01}\alpha_g+\frac{1}{2}h_{10}(5\alpha_g-3)  \right\}, \nonumber\\
A(u,\mu)&=&6u(1-u)\left\{
\frac{16}{15}+\frac{24}{35}a_2+20\eta_3+\frac{20}{9}\eta_4 \right.
\nonumber \\
&&+\left[
-\frac{1}{15}+\frac{1}{16}-\frac{7}{27}\eta_3\omega_3-\frac{10}{27}\eta_4\right]C^{\frac{3}{2}}_2(\xi)
\nonumber\\
&&\left.+\left[
-\frac{11}{210}a_2-\frac{4}{135}\eta_3\omega_3\right]C^{\frac{3}{2}}_4(\xi)\right\}+\left\{
 -\frac{18}{5}a_2+21\eta_4\omega_4\right\} \nonumber\\
 && \left\{2u^3(10-15u+6u^2) \log u+2\bar{u}^3(10-15\bar{u}+6\bar{u}^2) \log \bar{u}
 \right. \nonumber\\
 &&\left. +u\bar{u}(2+13u\bar{u})\right\} \, ,\nonumber\\
 g(u,\mu)&=&1+g_2C^{\frac{1}{2}}_2(\xi)+g_4C^{\frac{1}{2}}_4(\xi)\, ,\nonumber\\
 B(u,\mu)&=&g(u,\mu)-\phi(u,\mu)\, ,
\end{eqnarray}
where
\begin{eqnarray}
h_{00}&=&v_{00}=-\frac{\eta_4}{3} \, ,\nonumber\\
a_{10}&=&\frac{21}{8}\eta_4 \omega_4-\frac{9}{20}a_2 \, ,\nonumber\\
v_{10}&=&\frac{21}{8}\eta_4 \omega_4 \, ,\nonumber\\
h_{01}&=&\frac{7}{4}\eta_4\omega_4-\frac{3}{20}a_2 \, ,\nonumber\\
h_{10}&=&\frac{7}{2}\eta_4\omega_4+\frac{3}{20}a_2 \, ,\nonumber\\
g_2&=&1+\frac{18}{7}a_2+60\eta_3+\frac{20}{3}\eta_4 \, ,\nonumber\\
g_4&=&-\frac{9}{28}a_2-6\eta_3\omega_3 \, ,
\end{eqnarray}
 here $\xi=2u-1$, and $ C_2^{\frac{1}{2}}(\xi)$, $ C_4^{\frac{1}{2}}(\xi)$,
 $ C_1^{\frac{3}{2}}(\xi)$, $ C_2^{\frac{3}{2}}(\xi)$, $ C_4^{\frac{3}{2}}(\xi)$ are Gegenbauer polynomials,
  $\eta_3=\frac{f_{3K}}{f_K}\frac{m_u+m_s}{m_K^2}$ and  $\rho^2={(m_u+m_s)^2\over m_K^2}$
 \cite{LCSR,Belyaev94,PSLC}. The corresponding light-cone
 distribution amplitudes for the $\pi$ meson can be obtained with a
 simple replacement of the nonperturbative parameters.

\section*{Appendix \, C}
The light-cone distribution amplitudes of the $K^*$ meson are
defined
 by
\begin{eqnarray}
\langle 0| {\bar u} (0) \gamma_\mu s(x) |K^*(p)\rangle& =& p_\mu
f_{K^*} m_{K^*} \frac{\epsilon \cdot x}{p \cdot x} \int_0^1 du e^{-i
u p\cdot x} \left\{\phi_{\parallel}(u)+\frac{m_{K^*}^2x^2}{16}
A(u)\right\}\nonumber\\
&&+\left[ \epsilon_\mu-p_\mu \frac{\epsilon \cdot x}{p \cdot x}
\right]f_{K^*} m_{K^*}
\int_0^1 du  e^{-i u p \cdot x} g_{\perp}^{(v)}(u)  \nonumber\\
&&-\frac{1}{2}x_\mu \frac{\epsilon \cdot x}{(p \cdot x)^2} f_{K^*} m_{K^*}^3 \int_0^1 du e^{-iup \cdot x}C(u) \, ,\nonumber\\
 \langle 0| {\bar u} (0)  s(x) |{K^*}(p)\rangle  &=& \frac{i}{
2}\left[f_{K^*}^\perp-f_{K^*}
\frac{m_u+m_s}{m_{K^*}}\right]m_{K^*}^2\epsilon \cdot x  \int_0^1 du
e^{-i u p \cdot x} h_{\parallel}^{(s)}(u)  \,
,  \nonumber\\
\langle 0| {\bar u} (0) \sigma_{\mu \nu}  s(x) |{K^*}(p)\rangle
&=&i[\epsilon_\mu p_\nu-\epsilon_\nu p_\mu] f_{K^*}^\perp \int_0^1
du e^{-i u p \cdot x} \left\{\phi_{\perp}(u)+\frac{m_{K^*}^2x^2}{16}
A_{\perp}(u) \right\}   \nonumber\\
&&+i[p_\mu x_\nu-p_\nu x_\mu] f_{K^*}^\perp m_{K^*}^2\frac{\epsilon
\cdot x}{(p \cdot x)^2} \int_0^1 du e^{-i u p \cdot x}
 B_{\perp}(u)   \nonumber\\
 &&+i\frac{1}{2}[\epsilon_\mu x_\nu-\epsilon_\nu x_\mu] f_{K^*}^\perp m_{K^*}^2\frac{1}{p
\cdot x} \int_0^1 du e^{-i u p \cdot x}
 C_{\perp}(u)   \, ,\nonumber\\
\langle 0| {\bar u} (0) \gamma_\mu \gamma_5 s(x) |{K^*}(p)\rangle
&=& -\frac{1}{ 4}\left[f_{K^*}-f_{K^*}^\perp
\frac{m_u+m_s}{m_{K^*}}\right]m_{K^*}
\epsilon_{\mu\nu\alpha\beta}\epsilon^\nu p^\alpha x^\beta
\nonumber\\
&&\int_0^1 du e^{-i u p \cdot x} g_{\perp}^{(a)}(u)  \, .
\end{eqnarray}
The  light-cone distribution amplitudes of the $K^*$ meson are
parameterized as
\begin{eqnarray}
\phi_{\parallel}(u,\mu)&=&6u(1-u)
\left\{1+a_1^{\parallel}3\xi+a_2^{\parallel}
\frac{3}{2}(5\xi^2-1) \right\}\, , \nonumber\\
\phi_{\perp}(u,\mu)&=&6u(1-u) \left\{1+a_1^{\perp}3\xi+a_2^{\perp}
\frac{3}{2}(5\xi^2-1) \right\}\, , \nonumber\\
g_{\perp}^{(v)}(u,\mu)&=&\frac{3}{4}(1+\xi^2)+a_1^{\parallel}\frac{3}{2}\xi^3+\left\{ \frac{3}{7}a_2^{\parallel}+ 5\zeta_3^{\parallel}\right\}(3\xi^2-1)\nonumber \\
&&+\left\{ 5\kappa_3^{\parallel}-\frac{15}{16}\lambda_3^{\parallel}+\frac{15}{8}\widetilde{\lambda}_3^{\parallel}\right\}\xi(5\xi^2-3)\nonumber\\
&&+\left\{\frac{9}{112}a_2^{\parallel}+\frac{15}{32}\omega_3^{\parallel}
-\frac{15}{64}\widetilde{\omega}_3^{\parallel}\right\}(3-30\xi^2+35\xi^4)\,
,  \nonumber\\
g_{\perp}^{(a)}(u,\mu)&=& 6u\bar{u} \left\{1+\left(\frac{1}{3}a_1^{\parallel}+\frac{20}{9}\kappa_3^{\parallel} \right) C_1^{\frac{3}{2}}(\xi) + \right. \nonumber \\
&&\left.\left(\frac{1}{6}a_2^{\parallel}
+\frac{10}{9}\zeta_3^{\parallel}+\frac{5}{12}\omega_3^{\parallel}-\frac{5}{24}\widetilde{\omega}_3^{\parallel}\right)C_2^{\frac{3}{2}}(\xi)
+\left(
\frac{1}{4}\widetilde{\lambda}_3^\parallel-\frac{1}{8}\lambda_3^\parallel
\right)C_3^{\frac{3}{2}}(\xi) \right\} \nonumber\\
 h_{\parallel}^{(s)}(u,\mu)&=&6u\bar{u}
\left\{1+\left(\frac{a_1^\perp}{3}+\frac{5}{3}\kappa^{\perp}_3
\right)C_1^{\frac{3}{2}}(\xi)+\left(\frac{a_2^\perp}{6}+\frac{5}{18}\omega^{\perp}_3
\right)C_2^{\frac{3}{2}}(\xi)-\frac{1}{20}\lambda_3^{\perp}C_3^{\frac{3}{2}}(\xi) \right\}\, , \nonumber\\
h_{\parallel}^{(t)}(u,\mu)&=&3\xi^2+\frac{3}{2}a_1^{\perp}\xi(3\xi^2-1)+\frac{3}{2}a_2^{\perp}\xi^2(5\xi^2-3)+\frac{5}{8}
\omega_3^{\perp}(3-30\xi^2+35\xi^4)   \nonumber \\
&&+\left( \frac{15}{2}\kappa_3^{\perp}
-\frac{3}{4}\lambda_3^{\perp}\right)\xi(5\xi^2-3) \, ,\nonumber\\
g_3(u,\mu)&=&1+\left\{ -1-\frac{2}{7}a_2^{\parallel}+\frac{40}{3}\zeta_3^{\parallel} -\frac{20}{3}\zeta_4\right\}C_2^{\frac{1}{2}}(\xi)\nonumber \\
&&+\left\{-\frac{27}{28}a_2^{\parallel}
+\frac{5}{4}\zeta_3^{\parallel}
-\frac{15}{16}\widetilde{\omega}_3^{\parallel} -\frac{15}{8}\omega_3^{\parallel}\right\}C_4^{\frac{1}{2}}(\xi)\, ,  \nonumber \\
h_3(u,\mu)&=&1+\left\{ -1+\frac{3}{7}a_2^{\perp}-10(\zeta_4^T+\widetilde{\zeta}_4^T)\right\}C_2^{\frac{1}{2}}(\xi)+\left\{-\frac{3}{7}a_2^{\perp} -\frac{5}{4}\omega_3^{\perp}\right\}C_4^{\frac{1}{2}}(\xi)\, ,  \nonumber \\
A(u,\mu)&=&30u^2\bar{u}^2\left\{\frac{4}{5}
+\frac{4}{105}a_2^{\parallel}+\frac{8}{9}\zeta_3^{\parallel}+\frac{20}{9}\zeta_4\right\}\,
, \nonumber \\
A_{\perp}(u,\mu)&=&30u^2\bar{u}^2\left\{\frac{2}{5}
+\frac{4}{35}a_2^{\perp}+\frac{4}{3}\zeta_4^T-\frac{8}{3}\widetilde{\zeta}_4^T\right\}\,
, \nonumber \\
C(u,\mu)&=&g_3(u,\mu)+\phi_{\parallel}(u,\mu)-2g_{\perp}^{(v)}(u,\mu)
\, ,\nonumber \\
B_{\perp}(u,\mu)&=&h_{\parallel}^{(t)}(u,\mu)-\frac{1}{2}\phi_{\perp}(u,\mu)-\frac{1}{2}h_3(u,\mu)\, ,\nonumber \\
C_{\perp}(u,\mu)&=&h_3(u,\mu)-\phi_{\perp}(u,\mu) \, ,
\end{eqnarray}
where   $\xi=2u-1$, and  $C_2^{\frac{1}{2}}(\xi)$, $
C_4^{\frac{1}{2}}(\xi)$,
 $ C_1^{\frac{3}{2}}(\xi)$, $ C_2^{\frac{3}{2}}(\xi)$,  $
 C_3^{\frac{3}{2}}(\xi)$
   are Gegenbauer polynomials. The corresponding light-cone
 distribution amplitudes for the $\rho$ and $\phi$ mesons can be obtained with a
 simple replacement of the nonperturbative parameters.

\section*{Appendix \, D}
 Here we present some technical details
  necessary  in performing the Borel transformation which are not familiar to the
   novices,
\begin{eqnarray}
&&\int_0^1 dv \int_0^1 d\alpha_g \int_0^{1-\alpha_g} d\alpha_s
f(v,\alpha_s,\alpha_g) \frac{d}{du}
\exp\left[-\frac{m_c^2+u(1-u)m_K^2}{M^2}
\right]\nonumber \\
&&\delta(u-u_0)|_{u=\alpha_s+(1-v)\alpha_g} \nonumber \\
&=&\int_0^1 du\int_0^1 dv \int_0^1 d\alpha_g \int_0^{1-\alpha_g}
d\alpha_s f(v,\alpha_s,\alpha_g)
\delta\left[u-\alpha_s-(1-v)\alpha_g\right]\nonumber\\
&&\frac{d}{du} \exp\left[-\frac{m_c^2+u(1-u)m_K^2}{M^2}
\right]\delta(u-u_0) \nonumber \\
&=&\int_0^1 du \int_0^u d\alpha_s \int_{u-\alpha_s}^{1-\alpha_s}
d\alpha_g \frac{f(v,\alpha_s,\alpha_g)}{\alpha_g} \frac{d}{du}
\exp\left[-\frac{m_c^2+u(1-u)m_K^2}{M^2}
\right]\delta(u-u_0) \nonumber \\
&=&-\int_0^1 du \exp\left[-\frac{m_c^2+u(1-u)m_K^2}{M^2}
\right]\delta(u-u_0)\frac{d}{du}\int_0^u d\alpha_s
\int_{u-\alpha_s}^{1-\alpha_s} d\alpha_g
\frac{f(v,\alpha_s,\alpha_g)}{\alpha_g}
 \nonumber \\
 &=&- \exp\left[-\frac{m_c^2+u_0(1-u_0)m_K^2}{M^2}
\right]\frac{d}{du_0}\int_0^{u_0} d\alpha_s
\int_{u_0-\alpha_s}^{1-\alpha_s} d\alpha_g
\frac{f(v,\alpha_s,\alpha_g)}{\alpha_g} \, ,
 \nonumber
\end{eqnarray}
where the $f(v,\alpha_s,\alpha_g)$ stand for the three-particle
light-cone distribution amplitudes.

\section*{Acknowledgments}
This  work is supported by National Natural Science Foundation,
Grant Number 10405009,  and Key Program Foundation of NCEPU.

\end{document}